\documentclass[aps,prl,floatfix,twocolumn,showpacs]{revtex4}
\usepackage{latexsym}
\usepackage{graphicx}
\newcommand{\figwidth}{3.275 in}
\usepackage{epsfig}
\usepackage{amsmath}

\usepackage{color}
\begin{document}

\title{Limitations of the hybrid functional approach to electronic
structure   of transition metal oxides}
\author{John E. Coulter$^{1}$}
\author{ Efstratios Manousakis$^{(1,2)}$ }
\author{Adam Gali$^{(3,4)}$}\email{gali.adam@wigner.mta.hu}
\affiliation{
$^{(1)}$ Department  of  Physics and National High Magnetic Field Laboratory,
  Florida  State  University,  Tallahassee,  FL  32306-4350,  USA\\
$^{(2)}$Department   of    Physics,   University    of   Athens,
  Panepistimioupolis, Zografos, 157 84 Athens, Greece \\
$^{(3)}$Institute for Solid State Physics and Optics, Wigner
  Research Center for Physics,\\
  Hungarian Academy of Sciences, P.O.B. 49, H-1525, Budapest, Hungary \\
$^{(4)}$Department of Atomic Physics, Budapest
              University of Technology and Economics, Budafoki \'ut 8., H-1111,
              Budapest, Hungary}

\date{\today}

\begin{abstract}
During the last decade, \emph{ab initio} methods to calculate electronic
structure of materials based on hybrid functionals are increasingly becoming
widely popular. In this Letter, we show that, in the case of small gap
transition metal oxides, such as VO$_2$, with rather subtle
physics in the vicinity of the
Fermi-surface,  such hybrid functional
schemes without the inclusion of ``expensive'' fully self-consistent
GW corrections
fail to yield this physics and incorrectly describe the features of the wave
function of states near the Fermi-surface. While a fully self-consistent
GW on top of hybrid functional approach does correct
these wave functions as expected, and is found to be in general
agreement with the results of a fully self-consistent GW approach based
on semilocal functionals, it is much more computationally demanding
as compared to the latter approach for the benefit of essentially
the same results.
\end{abstract}
\maketitle

%

During the last few decades, new materials based on transition metal oxides (TMOs)
as the key parent component have surprised us with
their novel and unexpected behaviors. These include high temperature
superconductivity in the cuprates, giant magnetoresistance in
magnanites and
a plethora of fascinating new phenomena which has been recently reported on
oxide hetero-structures of TMOs and
devices \cite{Mannhart, Cen, Gozar}.
For example, an interface between
two insulators behaves as a metal \cite{Cen} which becomes
superconducting at sufficiently low temperatures, while an interface
between two antiferromagnets becomes ferromagnetic \cite{Millis}.
These new structures not only create a playground for unexpected
physical phenomena to be observed, but, in addition, they open up the
possibility for new applications based on radically different
foundations.   The complex, unusual, and as yet not fully
discovered or understood
behavior of small-gap TMOs can be manipulated in a variety of fundamentally
new applications \cite{Mottsolar}.

While in the late 80's, immediately after the discovery of the
cuprate superconductors, electronic structure calculations could
not predict the correct ground state of TMO based materials,
during the following decades significant progress has
been made which has restored some of the faith in the newly
developed \emph{ab initio} computational schemes.

The room temperature $M_1$ phase of VO$_2$, a
proto-typical  material\cite{Mott} in the  family of  TMOs, is  such a
small gap  system of correlated $d$-electrons.  Small-gap  TMOs have a
very rich, complex and  interesting phase diagram, where understanding
their electronic  structure and wave functions is  of high importance.
As a  result these  TMOs have been  widely studied  with sophisticated
\emph{ab initio} methods. A recent \emph{Letter} reported \cite{Eyert2}
that HSE06 non-local  range-separated hybrid density functional theory
(DFT) \cite{HSE} is  able to correctly describe the  ground state wave
functions of  VO$_2$, and produces a reasonable though  larger gap
than the  experimental one.  HSE  is a slower method  than traditional
(semi)local   DFT   functionals  but   much   faster  than   many-body
perturbation  methods, such  as the  GW-method or  the  dynamical mean
field  theory; thus, it  has been  assumed with  a growing  number of
followers
\cite{Heinemann-HSE, Zhuang-HSE, Gillen-HSE, Dutta-HSE}  that  HSE06 may
be  a very  practical method  to explore  the properties  of small-gap
TMOs.

%

Here  we   show  by  a  fully  dynamical   self-consistent  GW
(scGW) calculation  \cite{Hedin, Schilfgaarde}  that HSE06  does
not provide accurate wave functions and a semi-local DFT functional
is as good a starting  point for the  scGW calculation as  the
computationally demanding HSE06  method. Furthermore, we
demonstrate that even if the admixture  of  the Fock-exchange in
the hybrid  functional  is tuned  to reproduce the scGW
fundamental gap, still, the resulting wave functions deviate  from
those obtained by  scGW close  to the  Fermi-level.  We conclude
that (semi)local functionals  are as good starting points for the
scGW procedure as the hybrid functionals for small-gap TMOs.  The
former, however, is much less computationally demanding.


We  carried out  DFT  calculations on  the  $M_1$ phase  of VO$_2$  as
implemented  within  the  \textsc{VASP} package \cite{VASP, Kresse1,
  Kresse2}. We  used small  core projectors for  vanadium ions,  so we
explicitly included  $3s$ and $3p$  electrons as valence.  The valence
electron  states  were  expressed   as  linear  combinations  of  plane
waves.  We  found  that  the  plane wave  cutoff  of  400~eV  provided
convergent single particle levels. As we applied various functionals
and GW  schemes for calculating  the quasi-particle energies,  we used
the  experimental  geometry\cite{Anderson2}.
The Brillouin-zone integration was approximated
by a weighted sum on a special $k$-point set. We  found  that  the required
size  of  the Monkhorst-Pack  \cite{MP76}  $k$-point  set  depends
strongly  on  the existence  of  a gap. Convergent  charge
density could  be achieved with a $5\times5\times5$
Monkhorst-Pack $k$-point set
when there is a gap, while an $18\times18\times18$  $k$-point set was
required without a gap.  We applied 146 conduction bands in the
GW calculations.

The ground state electronic structure and wave functions were
calculated by the standard semi-local Perdew-Burke-Ernzerhof (PBE)
functional \cite{PBE} as well as non-local, range separated hybrid
functionals as proposed by Heyd-Scuseria-Ernzerhof (HSE)
\cite{HSE, HSE06}. The HSE functional for the exchange-correlation
part of the energy involves a parameter $\alpha$ which mixes the
contribution of the short-range parts of the Fock-exchange and the
PBE expression for the exchange energy \cite{HSE}. It also
involves a second parameter $\omega$ which defines what is meant
by the short and long ranged part of the Coulomb potential. 


The  value of  the parameter  $\omega=0.2  a_0^{-1}$ is determined
to  give a  balanced description that  provides good accuracy and
speed  for both molecules and    solids \cite{HSE06}. The
Fock-exchange part  is calculated using the  short-range part of
the  Coulomb  interaction.   The  PBE \cite{PBE}  expression for
the  exchange energy functional is modified   to  use   the short
and long   range  parts  of   the  Coulomb interaction \cite{HSE}.
The choice  of $\omega$ and $\alpha$ may depend on the actual
system.  We fixed the parameter $\omega$ at $0.2 a_0^{-1}$ while
we varied the parameter $\alpha$. This parameter is often tuned to
agree with   experimental data,   such   as  the band gap
\cite{Moussa-HSE, HSE-TunedLaMo, TunedHSE-Pozun}   or   the
dielectric constant \cite{Marques1} of  a  given   crystal.
$\alpha$=0.25 corresponds  to the HSE06 functional   \cite{HSE06}
that  we  call now HSE-0.250.   We note   that   $\alpha$=0.25 was
rationalized in Ref.~\onlinecite{PEB} where  they showed that a
smaller  value is needed for systems with nearly-degenerate
ground-states. VO$_2$ may fall into this  category,  thus  we
applied $\alpha$=0.125 and $\alpha$=0.172 (HSE-0.125 and
HSE-0.172 functionals, respectively).

As HSE functionals  contain external  parameters they  are  not truly
\emph{ab initio} methods. Nevertheless, HSE DFT functionals may provide
a good  starting  point  for  many-body perturbation  methods,  such  as
the GW-method  which may  result in quasi-particle  energies and
wave functions  that are ideally \emph{independent}  from the starting
point.   We applied  several levels  of approximations  within  the GW
quasi-particle   scheme   as   implemented  in   VASP \cite{Shishkin1,
  Shishkin2,   Fuchs,  Shishkin3}.  (i) First, we  applied   the  simplest
single-shot GW  approach, i.e., the G$_0$W$_0$  approximation.
This means that we have used the  Kohn-Sham eigenvalues and
orbitals in G$_0$ and W$_0$. For W we took W$_0=\epsilon^{-1}V$,
where the dielectric matrix $\epsilon^{-1}_{{\bf G},{\bf G'}}({\bf
q},\omega)$, with ${\bf G}$ and ${\bf G}'$  denoting reciprocal
lattice vectors, were  calculated in the random  phase
approximation   and  the  self-energy  corrections  were evaluated
to first order  in the  difference between  the self-energy
$\Sigma$ and  the Kohn-Sham potential \cite{Louie, Sham}. (ii) As
a higher level  approximation,  we solved for  G self-consistently
within the  GW approximation     following the     procedure
described     in Ref.~\onlinecite{Shishkin3}. Typically,  four
iterations  in  G  were sufficient  to achieve convergence of 
the self-consistent
quasi-particle  energies within  0.02~eV, and we denote this
procedure  by G$_4$W$_0$.  (iii) Last, we applied a fully
self-consistent  dynamical GW correction as proposed  by
van~Schilfgaarde \emph{et  al.} \cite{Schilfgaarde} and
implemented in  VASP \cite{Shishkin3}. In  this case, the  G and W
are updated  \emph{together}  with  the  wave  functions by means of 
the following equation:
\begin{equation}
\label{eq:scgw}
\left[T + V + \Sigma(E_n)\right] |\psi_n\rangle = E_n |\psi_n\rangle ,
\end{equation}
where $T$ is the kinetic energy operator, $V$ is the electrostatic
potential, and, $\Sigma(E_n)$ is the energy-dependent Hermitian
part of the self-energy as calculated from the GW approximation.
$E_n$ and $\psi_n$ are the quasi-particle energies and wave
functions. Formally, Eq.~\ref{eq:scgw} looks like an ordinary
Kohn-Sham DFT equation. Thus, the resulting wave
functions and energies can be used to recalculate G and W of the
system, and, the (Hermitian) part of $\Sigma(E_n)$, until
self-consistency has been reached. We found that typically 13-18
iterations were sufficient to obtain self-consistent
quasi-particle energies \emph{and} wave functions.
%
\begin{table}
\begin{ruledtabular}
\caption{We list the direct gaps in the $M_1$ phase of VO$_2$
         calculated using different methods and compare with the 
         results of other
         work and experimental values. Abbreviations of functionals
         are explained in the text. We note that
         PBE results in a metallic state (no gap).}
\begin{tabular}{ l r}
\label{tab:gaps}
 Source                         &   Gap          \\ \hline
 PBE (GGA)                       &   N/A       \\
 HSE-0.250                       &   1.01 eV       \\
 HSE-0.125                       &   0.26 eV        \\
 HSE-0.172                       &   0.55 eV        \\
 HSE-0.250+G$_0$W$_0$                   &   1.01 eV        \\
 HSE-0.125+G$_0$W$_0$                   &   0.69 eV        \\
 HSE-0.250+G$_4$W$_0$                   &   1.01 eV        \\
 HSE-0.125+G$_4$W$_0$                   &   0.69 eV        \\
 HSE-0.250+scGW                     &   0.54 eV        \\
 HSE-0.125+scGW                     &   0.54 eV        \\
 PBE+scGW                           &   0.54 eV        \\
 LDA+COHSEX+G$_0$W$_0$\footnotemark[1]        &   0.6  eV        \\
 LDA+empirical correction+G$_0$W$_0$\footnotemark[2] &   0.6  eV        \\
 Experiment\footnotemark[3]     &    $\sim$0.6 eV
\end{tabular}
\end{ruledtabular}
\footnotetext[1]{Ref.~\onlinecite{COHSEX} (with LDA lattice constant)}
\footnotetext[2]{Ref.~\onlinecite{LanyOxides} (empirical correction on d-orbitals)}
\footnotetext[3]{Ref.~\onlinecite{Epsilon2data}}
\end{table}

First, we discuss our results for the quasi-particle energies  close to
the  Fermi-level obtained with
 DFT and quasi-particle correction calculations
on the $M_1$ phase of VO$_2$.
The  $M_1$ phase  of  VO$_2$  has  a small  gap  of
$\sim$0.6~eV \cite{Epsilon2data}. PBE falsely predicts a metallic state
while HSE-0.250 (HSE06) yields too large a gap of 1.01~eV, in agreement
with a previous work \cite{Eyert2}. If  we apply HSE-0.125 then the gap
becomes too low at 0.26~eV. One  may assume that  HSE functionals are a
much better starting point for GW-calculation as they provide
a  gap, thus  G$_0$W$_0$ may  result  in good  results on  top of  HSE
functionals.  However,  G$_0$W$_0$  did  not improve  the  results  on
HSE-0.250.   The    calculated   gap   did   not   change,
(Table~\ref{tab:gaps})  which  might imply  that
HSE-0.250  produces   very  good  quasi-particle   energies  and  wave
functions.  However, the  G$_0$W$_0$  correction on  top of  HSE-0.125
gave   a   very   different   result, yielding a gap of   0.69~eV .  
When G was
self-consistently updated,  the quasi-particle energies 
did not change (see HSE-0.250+G$_4$W$_0$  and  
HSE-0.125+G$_4$W$_0$ results  in Table~\ref{tab:gaps}).
We conclude that G$_0$W$_0$ and G$_4$W$_0$ corrections do not supply a
ground state in close agreement with experiment near the Fermi-level. A
more complete calculation is needed to approach the experimental situation.

\begin{figure}
\epsfig{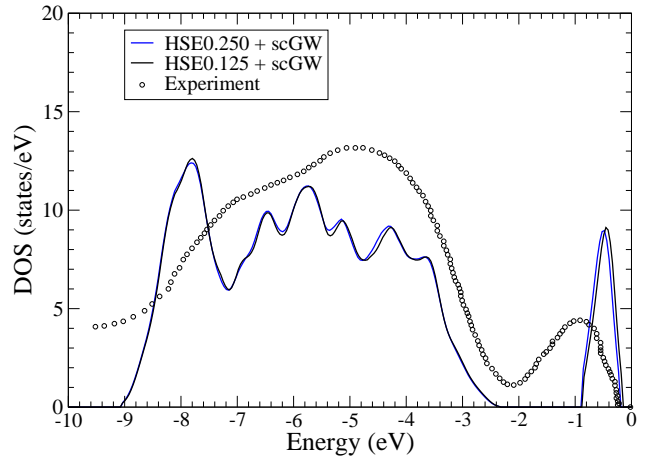}
\caption{(Color online) The photo-emission data taken from Ref.~\onlinecite{PES} is
  compared with the density of states (DOS) of the occupied states as calculated 
  within scGW on top of different HSE functionals. The calculated data was smoothed using an exponentially weighted moving average
  with smoothing factor of 0.25. We averaged each data point with its seven
neighbors using a decreasing weight $(1-\alpha)^n$, where $n$ is the number of 
points to the central point.}
\label{fig:PES}
\end{figure}

We then applied a  fully dynamical self-consistent GW-method where the
wave functions were updated together with the G and W. The calculated
HSE-0.250+scGW  and HSE-0.125+scGW  band gaps  are the  same
(0.54~eV), which  is  quite close  to  the  experimental  one. In
addition,  the calculated  density  of  states   agrees  well  with  the
experimental photo-emission spectrum (see Fig.~\ref{fig:PES}) \cite{PES}.

A       previous      theoretical      study       indicated  that the
local-density-approximation  (LDA) can  be a  good starting  point for the
self-consistent  GW procedure for  VO$_2$ \cite{COHSEX}. Gatti~\emph{et
  al.} applied the self-consistent    GW-method in the static   COHSEX
approximation  first,  which  opened  a  band gap  from  the
metallic  solution. Then,  they applied  a fully  dynamical G$_0$W$_0$
correction  on  the  quasi-particle  energies \cite{COHSEX}.  Here,  we
applied a fully dynamical self-consistent GW on top of a semi-local PBE
functional. The calculated PBE+scGW band gap  agrees well
with HSE+scGW. From this, it is fairly clear that starting from the
computationally ``expensive'' hybrid
functional  wave   functions  is   not  advantageous   when  a
self-consistent GW calculation is required,  as seems to be the case 
in many small-gap    systems,    such    as    the    TMOs \cite{LanyOxides}. 

\begin{figure*}
\includegraphics[keepaspectratio,totalheight=5 in, width= 6 in]{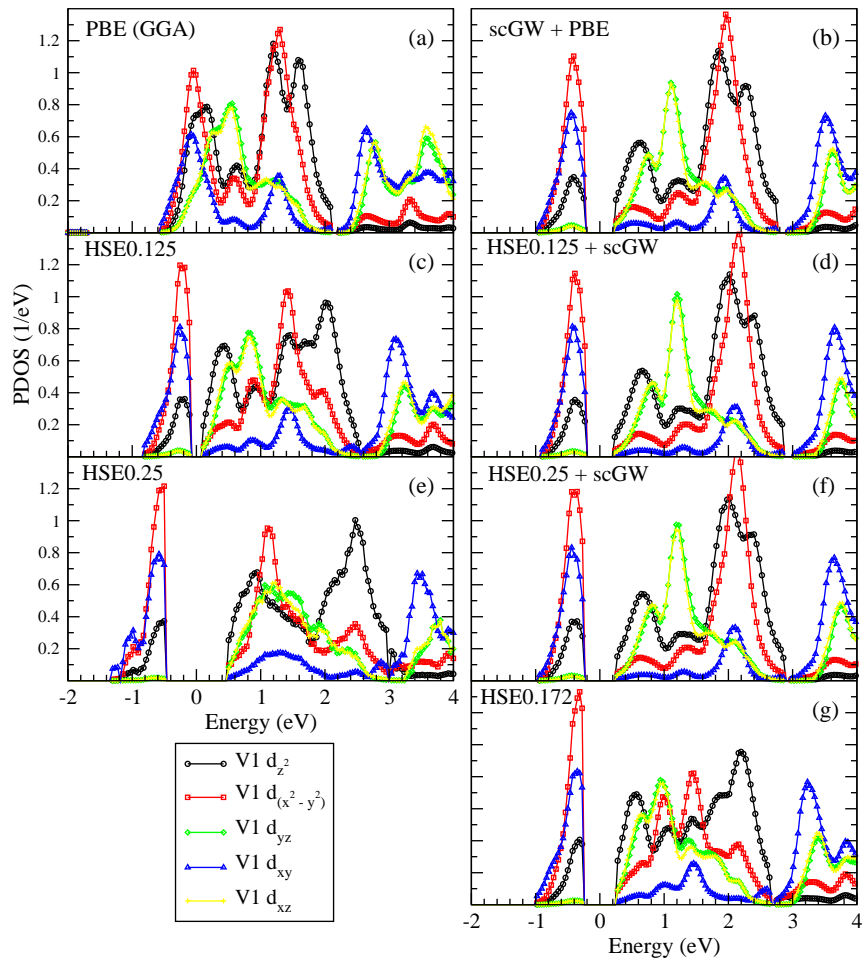}
\caption{(Color online) The partial density of states (PDOS) near the 
Fermi-level projected onto the 
spherical harmonics around one of
the unique vanadium atoms (V1) in the $M_1$ phase of VO$_2$, for a variety of functionals. 
The $\vec{x}$ axis is parallel to the $M_1$ "a" axis, 
$\vec{y}$ is parallel to the $M_1$ "b" axis \cite{Anderson2}, 
and $\vec{z}$ is  perpendicular to those, using a right-hand rule.} 
\label{fig:wfs}
\end{figure*}

As the wave functions are updated in the scGW procedure, it is intriguing
to study the change in the wave functions due to the scGW correction. This
analysis was already carried out when starting from LDA wave
functions \cite{COHSEX}.  Now, we  analyze   the  case  of  HSE
functionals. To  show the  change  in wave  functions,  we present  the
projected density of states (PDOS) onto the spherical harmonics around 
one unique vanadium atom close to the Fermi-level where the change is the most
significant. From these plots we  observed several interesting issues: (i) starting
with PBE  [c.f., Figs.~\ref{fig:wfs}(a) and (b)], the scGW procedure is  needed to open
the gap near  the Fermi-level so that the  contribution of d$_z^2$ and
(d$_{xz}$,d$_{yz}$)  will  be   significantly  smaller  than  that  of
d$_{xy}$ and  d$_{x^2-y^2}$ just below the  Fermi-level and \emph{vice
  versa}  just  above  the  Fermi-level, (ii)  the
unoccupied  wave  functions  with  energies of  $\sim$1~eV  above  the
Fermi-level in  PBE and PBE+scGW calculations are very similar, (iii)
HSE  naturally opens the gap [Figs.~\ref{fig:wfs}(c,e)],  but the
wave functions significantly differ from those
obtained with HSE+scGW
[c.f., Figs.~\ref{fig:wfs}(d) and (f)],  particularly,   at  energies  above  the
Fermi-level where  HSE+scGW yields similar  contributions from d$_z^2$
and d$_{x^2-y^2}$ orbitals at around 2~eV above the Fermi-level while
they      ``split''      in  the HSE     calculations. Apparently
[Figs.~\ref{fig:wfs}(b,d,f)],  the convergent  scGW wave  functions are
the same  \emph{regardless} of the  starting point. The  semi-local PBE
functional  provides relatively  good  wave functions for states
with energy 1~eV above  the
Fermi-level, \emph{unlike}  HSE-0.125 or HSE-0.250 functionals.

Recently,  it has  been claimed  that  for this  material, the  hybrid
functional  HSE-0.250 gives a  good description  of the
ground  state  \cite{Eyert2}.  According  to  our  analysis,  this  is
questionable. Apart  from close vicinity to the  Fermi-level, the
semi-local  PBE wave  functions are  superior to  the  HSE-0.250 wave
functions in the VO$_2$ crystal. Given  the similar number  of iterations
required to  achieve convergence in the scGW procedure,  and the decreased
workload,  it seems  that using  the HSE  family of  functionals  as a
starting  point for  the  more accurate  scGW  approximation does  not
provide an improved calculation at all.

We further  note that HSE  functionals have been  used as a  basis for
investigating previously  unknown materials \cite{Zhuang-HSE},  and to
explore the  complicated     physics     of      phase     transitions
\cite{Dutta-HSE,Heinemann-HSE}.  In addition, the parameter
$\alpha$ has been tuned to agree with experimental data such as the band gap
\cite{Moussa-HSE, HSE-TunedLaMo, TunedHSE-Pozun, Marques1}. 
In order to demonstrate the danger of fitting  the $\alpha$ parameter
to the band gap of $M_1$  phase of VO$_2$, we show that the ``tuned'' 
HSE functional does not produce the appropriate quasi-particle 
energies and wave functions. We
found, by a simple linear interpolation of our HSE-0.250 
and HSE-0.125 results, that  the HSE functional 
with $\alpha$=0.172 yields about the same gap of
0.55~eV as that found by the scGW approach. Comparing
the    wave     functions    of    scGW     and    HSE-0.172    [c.f.,
  Figs.~\ref{fig:wfs}(f) and (g)], it  is clear that the tuned functional and scGW calculations
significantly disagree  for the wave functions with energies above  the
Fermi-level.  Since  the  character  of  the wave  functions  is  quite
important   for  several   properties  such   as   optical  excitations
\cite{Bechstedt-New,  Schleife2011, Mottsolar},  it  seems that  using
such tuned functionals  as a starting point  may be not appropriate
for VO$_2$, and this ``tuning'' has to be carefully checked in 
small-gap TMOs.


In conclusion, we demonstrated that 
the HSE-type of functionals
should be  applied with  great care on  small-gap TMOs
such as  the $M_1$ phase  of VO$_2$. We  found that
scGW  calculations  on top  of  semi-local  PBE  or the  non-local  HSE
functionals  provide  the  same  results  and they are 
in agreement with  the  experimental
data. Furthermore, we have found that  the PBE wave functions  
for states which have energy 1~eV  above the
Fermi-level are superior, thus, 
the PBE provides at  least as good starting point as the HSE
functionals for detailed scGW calculations.


This work was supported in part by the U.S.\ National High Magnetic
Field Laboratory which is partially funded by the U.S.\ National
Science Foundation.

\end{document}